# Synthesis of endohedral iron-fullerenes by ion implantation[a]

H. Minezaki,[1] S. Ishihara,[1] T. Uchida,[2,b] M. Muramatsu,[3] R. Rácz,[4] T. Asaji,[5] A. Kitagawa,[3] Y. Kato,[6] S. Biri,[4] and Y. Yoshida[1,2]

[1]*Graduate School of Engineering, Toyo University, 2100, Kujirai, Kawagoe, Saitama 350-8585, Japan*  [2]*Bio-Nano Electronics Research Centre, Toyo University, 2100, Kujirai, Kawagoe, Saitama 350-8585, Japan*

[3]*National Institute of Radiological Sciences (NIRS), 4-9-1, Anagawa, Inage-ku, Chiba-shi, Chiba 263-8555, Japan*

[4]*Institute of Nuclear Research (ATOMKI), Bem tér 18/C, H-4026 Debrecen, Hungary*

[5]*Oshima National College of Maritime Technology, 1091-1, Komatsu Suou Oshima-city Oshima, Yamaguchi 742-2193, Japan*

[6]*Graduate School of Engineering, Osaka University, 2-1, Yamada-oka, Suita-shi, Osaka 565-0871, Japan*

In this paper, we discuss the results of our study of the synthesis of endohedral iron-fullerenes. A low energy $Fe^+$ ion beam was irradiated to $C_{60}$ thin film by using a deceleration system. $Fe^+$-irradiated $C_{60}$ thin film was analyzed by high performance liquid chromatography and laser desorption/ionization time-of-flight mass spectrometry. We investigated the performance of the deceleration system for using a $Fe^+$ beam with low energy. In addition, we attempted to isolate the synthesized material from a $Fe^+$-irradiated $C_{60}$ thin film by high performance liquid chromatography.

## I. INTRODUCTION

We have developed an electron cyclotron resonance ion source (ECRIS) apparatus, which is designed for endohedral fullerene production.[1–3] Endohedral fullerenes have at least one additional atom within a fullerene cage. They exhibit new physical and chemical characteristics by involving an atom or a molecule. Their production can be realized through the use of arc discharge, laser vaporization, and plasma. Among those methods, we focused on the method of using the ECRIS plasma.

There are two ways to produce endohedral fullerenes by an ECRIS due to collision reaction of ion-ion or neutralion in the ECRIS plasma or by ion implantation. Using the ECRIS plasma, our authors reported endohedral nitrogen-$C_{60}$ (N@$C_{60}$) production in the mixture plasma of nitrogen and $C_{60}$.[4] Using ion implantation, Watanabe *et al.* have reported endohedral Xe-$C_{60}$ (Xe@$C_{60}$) production by Xe ion implantation to a $C_{60}$ thin film from 30 to 38 keV.[5]

Recently we have synthesized Fe-$C_{60}$ complex by ion implantation.[6] Our aim is to produce endohedral iron-$C_{60}$ (Fe@$C_{60}$), which could be applied as a contrast material for magnetic resonance imaging. In the study noted above, we developed a deceleration system in order to irradiate the single-charged iron ion ($Fe^+$) beam to $C_{60}$ thin film at low energy. We then irradiated $Fe^+$ ion beam to $C_{60}$ thin film samples, and we were able to observe a Fe+$C_{60}$ peak with a time-of-flight mass spectrometer. Additionally, in order to investigate the structure of the Fe+$C_{60}$, we separated the synthesized material of the Fe+$C_{60}$ from the $Fe^+$-irradiated $C_{60}$ thin film samples using high performance liquid chromatography (HPLC). We verified that the shape and the size of the Fe+$C_{60}$ are similar to the $C_{60}$.

In the present paper, in order to investigate the performance of the deceleration system, the current, which flowed into each electrode, was measured. Additionally, we tried to isolate the Fe+$C_{60}$ from the $Fe^+$-irradiated $C_{60}$ thin film by HPLC. In general, isolation of the endohedral $C_{60}$ is very difficult. Jakes *et al.* have succeeded in the separation of N@$C_{60}$ by recycling the eluted fraction in a HPLC chromatogram.[7] Therefore, we recycled the HPLC procedure using the eluted fraction in the previous HPLC of $Fe^+$-irradiated $C_{60}$ thin film.

## II. EXPERIMENTAL

### A. Characterization of the deceleration system

In this experiment, the deceleration system of the Bio-Nano ECRIS was used.[6] Fig. 1 shows a schematic drawing of the deceleration system, which consists of a beam restriction electrode, a suppressor electrode, deceleration electrodes I, II, and a substrate holder. The beam restriction electrode restricts the ion beam by a $\phi$20 mm-aperture. The deceleration electrodes I, II, and substrate holder decelerate the ion beam, extracted at high voltage, by being applied a positive voltage. The potential of the suppressor electrode is 200 V lower than that of the deceleration electrodes I, II, and substrate holder. Therefore, the suppressor electrode can repel secondary electrons emitted from the substrate by ion beam irradiation. In this experiment, in order to calculate the exact dose, the aperture of deceleration electrode I was changed from $\phi$20 to $\phi$6 mm. The irradiation area of the ion beam could be identified. Additionally, when the $Fe^+$ beam was decelerated, the current, which flowed into each electrode, could be measured. The $Fe^+$ beam was generated in the ECRIS using ferrocene ($C_{10}H_{10}Fe$, Wako

---

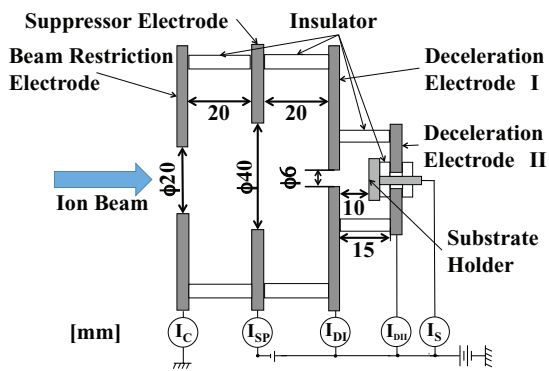

FIG. 1. Schematic drawing of the deceleration system.

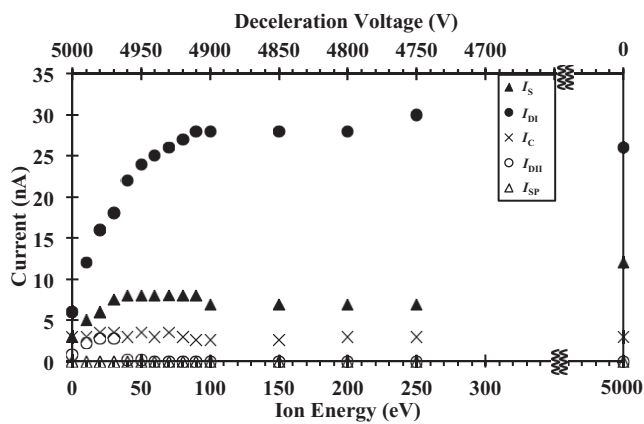

FIG. 2. Deceleration voltage (ion energy) dependence of the current to each electrode. The current, which flowed into each electrode, was written as follows; substrate holder: $I_S$, deceleration electrode I: $I_{DI}$, deceleration electrode II: $I_{DII}$, suppressor electrode: $I_{SP}$, and beam restriction electrode: $I_C$.

source. The experimental conditions were as follows: the extraction voltage was 5.0 kV, the deceleration voltage ranged from 4.75 to 5.0 kV (the ion energy from 250 to 0 eV), and the Fe$^+$ beam current was 40 nA.

### B. Preparation and analysis of the Fe$^+$-irradiated C$_{60}$ thin film

In the generation of the Fe$^+$-irradiated C$_{60}$ thin film, a pre-prepared 10 nm-thick C$_{60}$ thin film (size of 5 × 5 mm$^2$) was set up on the substrate holder. The Fe$^+$ beam was extracted from the ECRIS forming the ion energy of 5 keV, and decelerated in the ion energy of 50 eV by using the deceleration system. The other irradiation conditions of the Fe$^+$ beam
are as follows: the Fe$^+$ beam current of 20 nA (the current at the substrate holder), and an irradiation time of 75 s. The dose of Fe$^+$, which was calculated by the beam size ($\phi$6 mm) and the irradiation time, was 3.3 × 10$^{13}$ ions/cm$^2$.

The parameters of the ECRIS for production of the Fe$^+$ beam are as follows: microwave frequency of 9.75 GHz, microwave power ranged from 10 to 20 W, mirror coils currents ranged from 424 to 500 A (maximum axial magnetic field strength ranged from 0.37 to 0.44 T), ferrocene heating temperature of room temperature, and pressure of 1.0 × 10$^{-5}$ Pa, extraction voltage of 5.0 kV.

HPLC analyses were performed using the Prominence HPLC system (Shimadzu Corporation). The HPLC data were taken with a Buckyprep M (250 mm × 4.6 I.D. mm, Nacalai Tesque, INC.) column with the mixture of toluene/hexane 80/20 v/v as a mobile phase. The flow rate of the mobile phase was 0.7 ml/min. The 120 samples of the Fe$^+$-irradiated C$_{60}$ thin film were dissolved in solvent, which is the same as the mobile phase, with the concentration of 13 $\mu$g/ml. The injected sample solution volume was 0.02 ml. The HPLC chromatogram was measured by optical absorption at 335 nm. In this HPLC analysis, fullerenes were separated by molecular geometry.[8,9] The eluted fraction was collected and analyzed again using a small portion of the eluted fraction by the LDI-TOF-MS (Autoflex II, Bruker Daltonics Inc.). Then, we repeated the above HPLC/LDI-TOF-MS procedure using the eluted fraction several times: the eluted fraction was recycled

### III. RESULTS AND DISCUSSION

#### A. Characterization of the deceleration system

Fig. 2 shows the deceleration voltage (ion energy) dependence of the current to each electrode. The current, which flowed into the substrate holder (hereinafter, $I_S$), was constant within the ion energy range from 250 to 40 eV and decreased within the ion energy range from 30 to 0 eV. The current, which flowed into the deceleration electrode I (hereinafter, $I_{DI}$), was constant within the ion energy range from 250 to 80 eV, and decreased within the ion energy range from 70 to 0 eV. The current, which flowed into the deceleration electrode II (hereinafter, $I_{DII}$), was 0 nA within the ion energy range from 250 to 40 eV and was approx. 3 nA within the ion
energy range from 30 to 0 eV. The current, which flowed into the suppressor electrode (hereinafter, $I_{SP}$), was 0 nA within the ion energy range from 250 to 0 eV. The current, which flowed into the restriction electrode (hereinafter, $I_C$), was constant within the ion energy range from 250 to 0 eV. Compared with the $I_S$ and $I_{DII}$, $I_S$ decreased within the ion energy range from 30 to 0 eV, while positive $I_{DII}$ was observed within the ion energy range from 30 to 0 eV. These results indicate that the ion beam did not diverge within the ion energy range from 250 to 40 eV, and diverged within the ion energy range from 30 to 0 eV. We irradiated the Fe$^+$ beam within the ion energy range from 250 to 40 eV in our previous paper.[6] Therefore, we could confirm the previous data integrity is high from these results.

#### B. Analysis of the Fe$^+$-irradiated C$_{60}$ thin film by HPLC and LDI-TOF-MS

Fig. 3 shows the HPLC chromatogram of the Fe$^+$-irradiated C$_{60}$ thin film. We observed the peaks at retention time of 14.1 min and 16.2 min in the chromatogram. The 14.1 min and the 16.2 min retention time correspond to the C$_{60}$ and to the O+C$_{60}$ molecules respectively. In our previous study, we observed the Fe+C$_{60}$ peak in the C$_{60}$ one. According to the previously published results, endohedral X@C$_{60}$ peaks (X = Ar, Eu, N) were observed at near the C$_{60}$ one in the HPLC chromatogram.[10–12] In addition, for the separation of the endohedral C$_{60}$ peaks from the C$_{60}$ one, it is necessary to recycle the separation of the first half and second half of the C$_{60}$ peak into different fractions several times.[7] Therefore, in order to investigate the position of the Fe+C$_{60}$, we also separated the first half and second half of the C$_{60}$ peak into different fractions, and they were analyzed by LDI-TOF-MS.

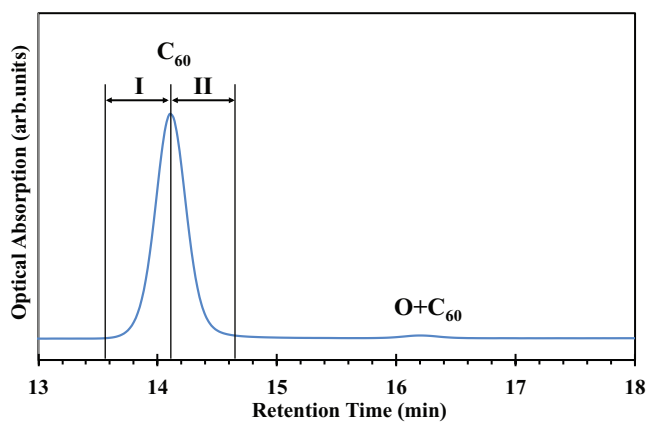

FIG. 3. HPLC chromatogram of the Fe$^+$-irradiated C$_{60}$ thin film.

Fig. 4 shows the LDI-TOF-MS spectrum of fraction II. We were able to observe the peak of the C$_{60}$ and also that of the Fe+C$_{60}$. Regarding fraction I, we could not observe the peak of the Fe+C$_{60}$ but that of the C$_{60}$. Therefore, we believe that the Fe+C$_{60}$ complex contained in fraction II. Then, we recycled the HPLC procedure using fraction II in the previous HPLC of Fe$^+$ irradiated C$_{60}$ thin film several times in order to separate Fe+C$_{60}$ from C$_{60}$.

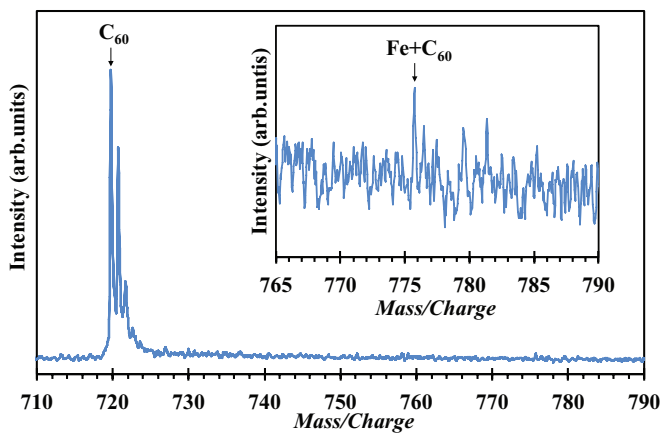

FIG. 4. LDI-TOF-MS spectrum of fraction II.

The peak intensity of the C$_{60}$ peak in the HPLC chromatogram decreased up to fourth recycling process. However the separation of the Fe+C$_{60}$ peak from the C$_{60}$ one could not be observed. It is supposed that the quantity of the Fe+C$_{60}$ is still insufficient. Therefore, in order to increase the quantity of the Fe+C$_{60}$, it is thought that a minor change in the deceleration system is required. For example, a structure such as the multi-sample is needed to be attached to the substrate holder without changing the deceleration voltage.

Nevertheless, we were able to observe the Fe+C$_{60}$ from the LDI-TOF-MS spectrum of the eluted fraction again and were able to affirm the reproducibility of the previous data.

## ACKNOWLEDGMENTS

This work was supported by JSPS KAKENHI Grant Nos. 24810029 and 24710095. The participation of two of the authors (S.B. and R.R.) in this work was partly supported by the TAMOP 4.2.2.A-11/1/KONV-2012-0036 project, which is co-financed by the European Union and European Social Fund. Part of this study has been supported by a Grant for the Strategic Development of Advanced Science and Technology S1101017 organized by the Ministry of Education, Culture, Sports, Science and Technology (MEXT) since April 2011.